# Directly created electrostatic micro-domains on hydroxyapatite: probing with a Kelvin Force probe and a protein

Tomas Plecenik[1,*], Sylvain Robin[2], Maros Gregor[1], Martin Truchly[1], Sidney Lang[3], Abbasi Gandhi[2], Miroslav Zahoran[1], Fathima Laffir[2], Tewfik Soulimane[2], Melinda Vargova[4], Gustav Plesch[4], Peter Kus[1], Andrej Plecenik[1] and S. A. M. Tofail[2]

[1]Department of Experimental Physics, Faculty of Mathematics, Physics and Informatics, Comenius University, 842 48 Bratislava, Slovakia
[2]Material and Surface Science Institute, University of Limerick, Ireland
[3]Department of Chemical Engineering, Ben-Gurion University of the Negev, Beer Sheva, Israel
[4]Department of Inorganic Chemistry, Faculty of Natural Sciences, Comenius University, 842 15 Bratislava, Slovakia

[*]corresponding author; e-mail: tomas.plecenik@fmph.uniba.sk ; phone: +421 2 60295517; fax: +421 2 60295867

**Abstract**
Micro-domains of modified surface potential (SP) were created on hydroxyapatite (HAp) films by direct patterning by mid-energy focused electron beam, typically available as a microprobe of Scanning Electron Microscopes. The SP distribution of these patterns has been studied on sub-micrometer scale by the Kelvin Probe Force Microscopy method as well as lysozyme adsorption. Since the lysozyme is positively charged at physiological pH, it allows us to track positively and negatively charged areas of the SP patterns. Distribution of the adsorbed proteins over the domains was in good agreement with the observed SP patterns.

## 1 Introduction

Properties of biomaterials surfaces are known to play significant role in many medical and biological applications, particularly in the field of medical implants. In particular, it is commonly understood that electrical properties such as local electrostatic charge distribution at biomaterials surface play a significant role in defining biological interactions such as protein or bacterial adsorption [1, 2]. However, the intertwined nature of such modifications poses a problem in determining the exact role of different processes in a biological environment. Various methods have been used to modify biomaterials surface properties in order to control its interaction with proteins and biological cells. Among others, these include photon irradiation [3], application of external electric field [4], plasma modification [4], ion beam irradiation [5] and electron beam irradiation [6, 7]. Furthermore, it has also been demonstrated that thin hydroxyapatite films are both piezoelectric and pyroelectric [8]. This may also play a role in biological interactions.

Recently, we have reported a convenient technique of direct microscopic patterning of surface potential (SP) on hydroxyapatite by a focused electron beam, typically available as an electron microprobe in a Scanning Electron Microscope (SEM) [7]. We have been successful to create circular, doughnut shaped, and bimodal micrometer-sized patterns with positive and negative SP. In the present study, we investigate into the influence of such micro-domains on protein placement. Using hen egg white Lysozyme (LSZ), which exhibits an overall positively charged surface at physiological pH, as a probing protein, we were able to track positively and negatively charged areas of the SP patterns. Our results show, that the protein adhesion is in a good agreement with the observed SP distribution of the domains. Moreover, by this experiment we have shown that the adhesion of proteins to biomaterials surfaces can be conveniently adjusted on at least micrometer scale by this method.

## 2 Experimental methods

Hydroxyapatite (HAp) thin films have been deposited on Si substrates by spin-coating through a sol-gel synthesis route. The film quality has been verified and confirmed as nano-crystalline HAp by x-ray diffraction (XRD: Philips X'Pert Pro) and x-ray photoelectron spectroscopy (XPS: Kratos Axis 165). The films thickness has been determined in cross-section by SEM (SEM: Jeol JSM-840) to be about 340 nm.

SEM TESCAN TS 5136 MM was used for the SP domain creation. A matrix of 4x4 points with approximately 18 μm separation has been irradiated on these HAp films by focused electron beam of the SEM with an absorbed beam current of 1.4 nA, a beam diameter of about 340 nm and beam exposure time of 70 seconds/exposure point at an electron energy of 20 keV. During the irradiation, the vacuum was kept at $10^{-4}$ Pa. No change of the surface topography caused by the irradiation was observed by AFM. Previously done XPS measurements on such irradiated areas also didn't show any significant changes of the HAp chemistry, although some slight carbon/hydrocarbon surface contamination took place [7].

Surface topography of the films was further inspected by Scanning Probe Microscope (NT-MDT Solver P47 PRO) in semicontact Atomic Force Microscopy (AFM) regime using standard silicon AFM tips. This SPM with TiN-coated conductive tips was also used for the SP measurements by Kelvin Probe Force Microscopy (KPFM).

SP of the irradiated area was than inspected by KPFM. To investigate the protein adsorption onto irradiated HAp film, fluorescein isothiocyanate (FITC)-labeled hen egg white Lysozyme was employed as a model protein. The LSZ is a small "hard" protein [9] of 14 kDa, as its fold is reinforced by the presence of 4 disulfide bridges, providing a strong internal coherence. The pI of LSZ is 11, therefore, at neutral pH, the polypeptide presents an overall positive surface charge [10]. Before protein immobilization, the FITC-LSZ solution was prepared with a concentration of 0.5 mg/ml in 1mM HEPES buffer at pH 7.3. After incubation for 1 min. at 20°C, the samples were rinsed with the same buffer and water. The samples were finally air-dried and the fluorescence was imaged by a laser scanning confocal microscope (Carl Zeiss Meta710) to determine the protein distribution.

The model figures were produced using a finite element program called FlexPDE (PDE Solutions, Inc., Antioch, CA). The figure shows an 80 by 80 micron area containing 16 charged regions. The charge is uniformly distributed over a 9 micron diameter area. The charged areas have an 18 micron center-to-center spacing. The voltage on each charged area is approximately -30 mV. The voltage at points distant from the charged areas is less than -0.2 mV.

## 3 Results and discussion

Typical distribution of the adsorbed lysozyme observed by confocal microscopy, KPFM image of the same irradiated area as well as corresponding modeled potential distribution is shown on Fig. 2. Single-domain comparison and its profile cross-section of both the protein adsorption (luminescence) and SP distribution are shown on Fig. 3.

As the ionic strength of the protein solution is low (1 mM HEPES), in the present conditions, the adsorption of LSZ onto HAp surface is mainly driven by electrostatic interactions [11]. As the LSZ is a "hard" protein, its adsorption is enthalpy-driven and does not involve any structural rearrangement in contrary to proteins with lower structural stability, such as, for instance, bovine serum albumin (BSA) [12]. For this reason, the adsorption of LSZ is only driven by the surface properties. The protein adsorbs onto the surface and presumably accumulates forming a loosely packed monolayer [13, 14] that has been found to neutralize the charges created during patterning [14].

Depending on the electron dose injected, the HAp hydrophobicity can be increased. In the present set up, it is unlikely that the change in wettability can explain the behavior of the protein used to probe the domains. Indeed, numerous studies have shown that the adsorption of LSZ increases with the hydrophobicity [15]. After irradiation, a higher LSZ adsorption would be expected. At pH 7.3, the LSZ is positively charged and at low ionic strength the electrostatic interactions play a major role in the binding of this protein to hydroxyapatite.

If we take a cross-section of a single domain, the distribution of both SP and protein attachment can be generally divided into four areas (Fig. 3). In the first area, the SP is increased in comparison to the domain surroundings and the protein adhesion is highly decreased or none. In contrary, in the second area the SP decreases back approximately to the surroundings level and the protein attachment considerably increases. Although the SP increases again in the third area, the protein adhesion remains relatively high. Finally, in the fourth area the SP decreases to the lowest value, while the protein adhesion increases to highest levels.

In general, we can say that the adsorption of the positively charged LSZ is favored onto the negatively charged areas (i.e. the areas with lower SP) and follows the SP distribution according to the expectations. This clearly shows that it is possible to use the protein adsorption to track positively and negatively charged areas of the biomaterials surface. However, although the lysozyme was clearly preferentially attached to negatively charged areas, in some parts there is a disproportion between the protein attachment and the SP. For example, areas 1 and 3 have very similar SP, but the protein adhesion was considerably higher in the 3$^{rd}$ area. We propose that this can be caused by the influence of an e-beam induced carbon/hydrocarbon contamination layer.

Our previously done XPS measurements on large array of similar domains showed that although there are no significant changes in the HAp chemistry, there is some evidence of carbon/hydrocarbon surface contamination induced during the e-beam irradiation [7]. Creation of such contamination layer during an electron beam irradiation was observed also by Aronov et al. on HAp films [6] as well as on other materials [16]. As this carbon/hydrocarbon layer may have significant influence on both SP and protein adhesion, it is very probable that it is in a large extent responsible for the observed disproportions between the protein adsorption and the SP distribution.

## 4 Conclusions

Micro-domains of modified surface potential have been created on hydroxyapatite films by focused electron beam irradiation. We have shown that by creation of such areas of modified surface potential, the adhesion of proteins to biomaterials surfaces can be adjusted on at least micrometer-scale. Our results also show that it is possible to use proteins with non-zero surface charge to track positively and negatively charged areas of biomaterials surfaces. Distribution of the adsorbed proteins was in a good agreement with the observed SP distribution of the domains. Some disproportions are explained by an influence of an e-beam induced carbon/hydrocarbon surface contamination layer. Presented findings may have important practical applications in the field of medical implants where adsorption of specific proteins need to be adjusted as well as within the biosensors and tissue engineering research and applications.

**Acknowledgments**
This project has been funded with support from the European Commission (EC NMP4-SL-2008-212533 – BioElectricSurface). This publication reflects the views only of the authors, and the Commission cannot be held responsible for any use which may be made of the information contained therein. This work was also supported by the Slovak Research and Development Agency under the contract Nos. DO7RP-007-09 and APVV-0199-10, by the Ministry of Education of the Slovak Republic under Contract Nos VEGA 1/0162/10 and VEGA 1/0605/12 and is also the result of the project implementation: 2622020004 supported by the Research & Development Operational Programme funded by the ERDF.

**Figures**

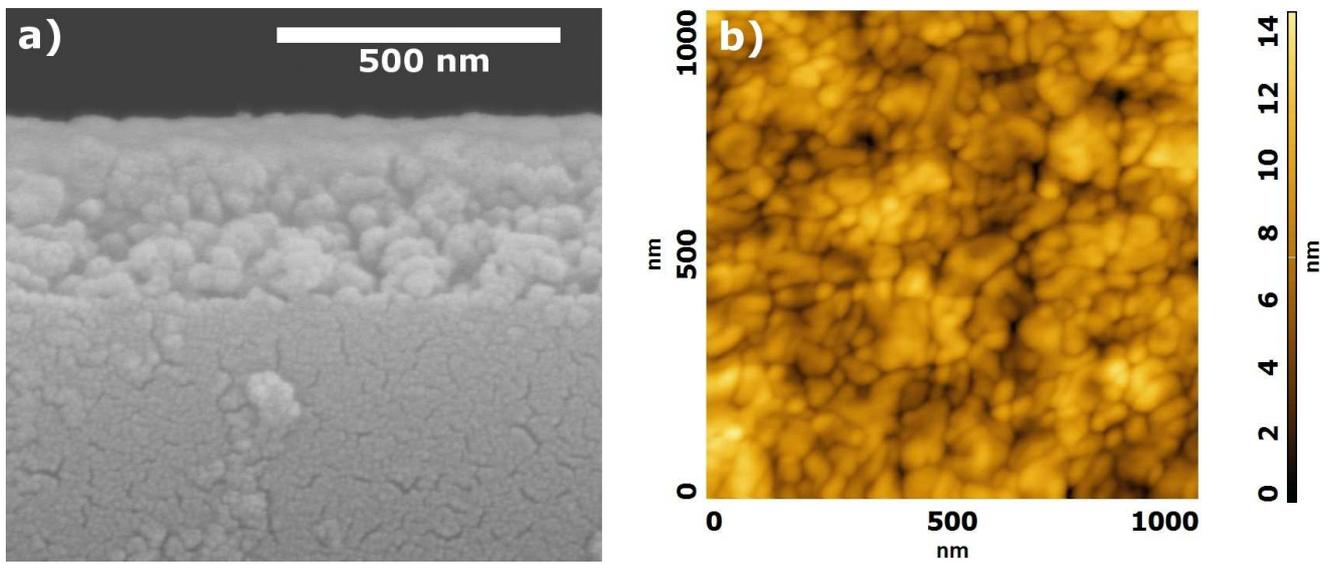

*Fig. 1: a) The HAp film cross-section obtained by SEM b) AFM surface topography of the film.*

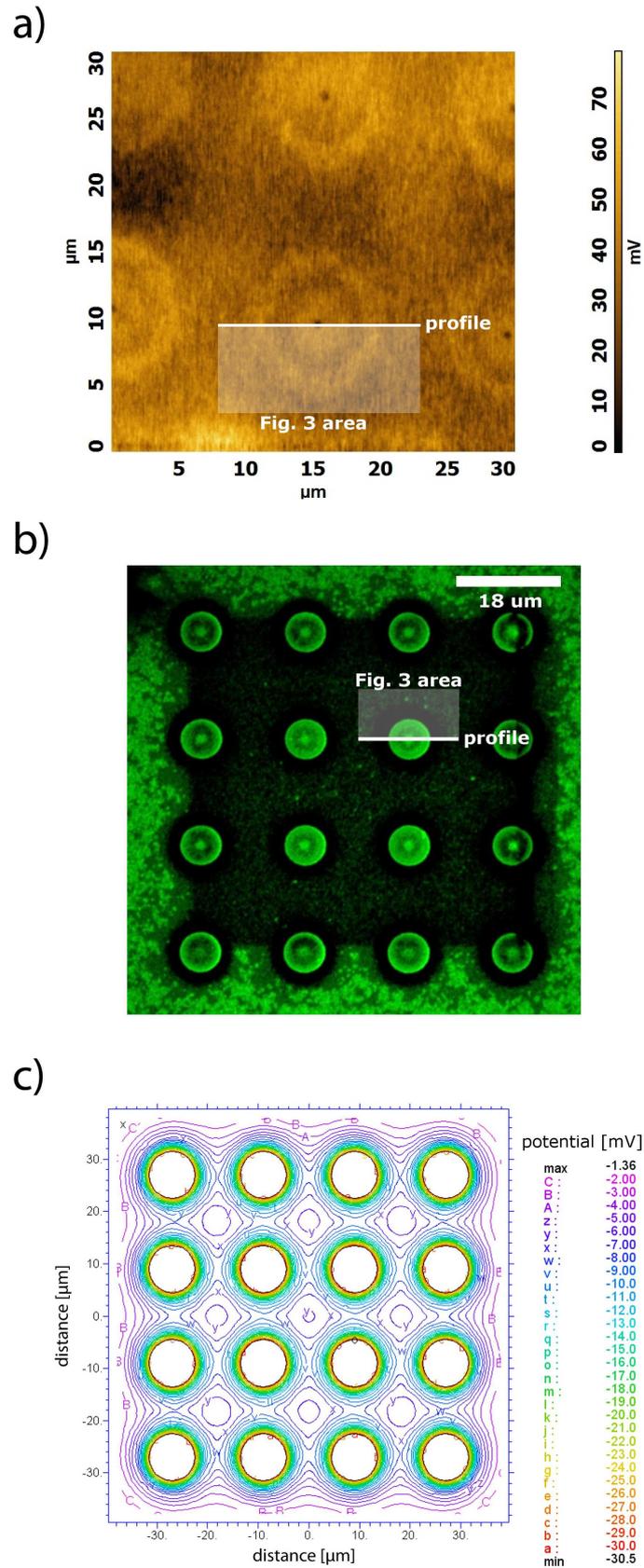

*Fig. 2: a) FTIC-Lysozyme adsorption (LSCM) at the array of irradiated domains. b) Relative surface potential distribution of the same domain array measured by KPFM c) modeled potential distribution of an array of 4x4 charged domains*

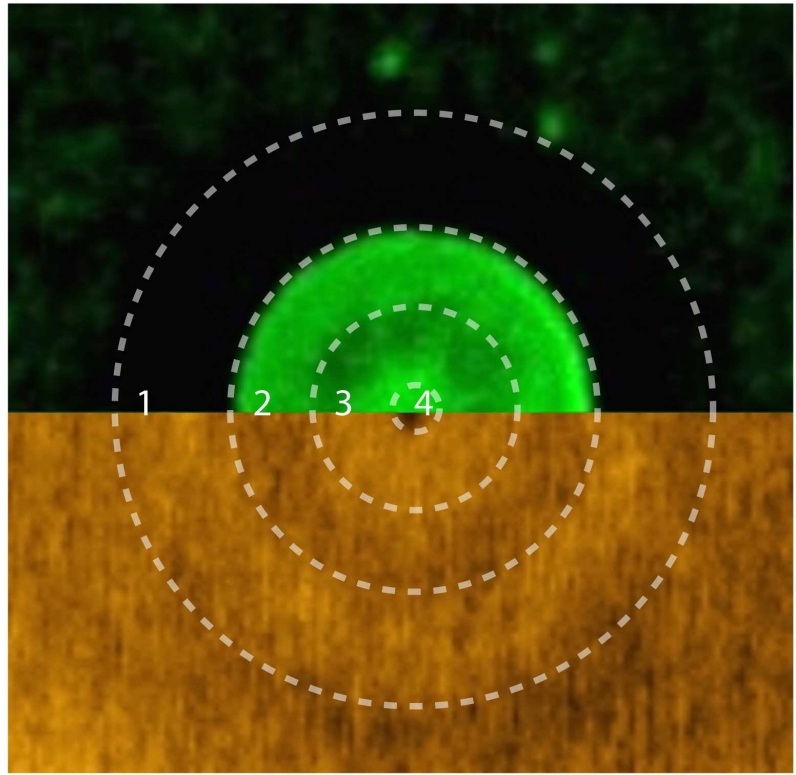

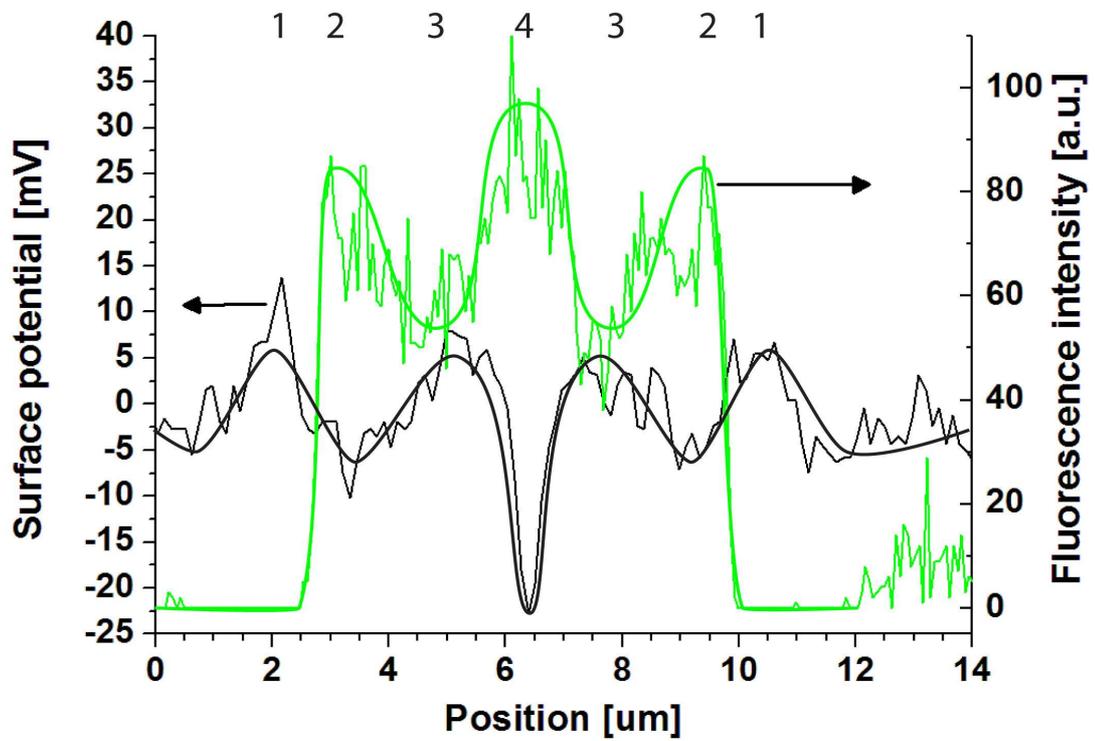

*Fig. 3: Profile of the protein adhesion (fluorescence) and relative SP distribution across the single domain shown on Fig. 2*